\documentclass[twocolumn,  twocolappendix]{aastex63}

\shorttitle{Hydroxylated Mg-rich Amorphous Silicates}
\shortauthors{Mennella et al.}

\begin{document}

\title{Hydroxylated Mg-rich amorphous silicates as catalysts for molecular hydrogen formation in the interstellar medium}

\correspondingauthor{Vito Mennella}
\email{vito.mennella@inaf.it }

\author[0000-0001-9525-895X]{V. Mennella}
\affiliation{INAF--Osservatorio Astronomico di Capodimonte, via Moiariello 16, 80131, Napoli, Italy \\}

\author{T. Suhasaria}
\affiliation{Max Planck Institute f\"ur Astronomie, K\"onigstuhl 17, 69117, Heidelberg, Germany
\\}

\author{B. Kerkeni}
\affiliation{ISAMM, Universit\'e de la Manouba, La Manouba 2010, Tunisia}
\affiliation{D\'epartement de Physique, LPMC, Facult\'e des Sciences de Tunis, Universit\'e de Tunis el Manar,Tunis 2092, Tunisia \\}
 
\author{G. Ouerfelli}
\affiliation{Department of Physics, College of Khurma University, Taif University, P.O. Box 11099, Taif 21944, Saudi Arabia
 \\}

\begin{abstract}
We present results from an experimental study on the interaction of atomic deuterium with Mg-rich amorphous enstatite and forsterite type silicates. Infrared spectroscopy was used to examine the process. During D atom exposure, deuteroxyl group formation was observed. The cross section for OD group formation, estimated from the OD stretching band intensity with D atom fluence, is $\sigma _{f,OD}$  = 4.2$\times10^{-18}$ cm$^{2}$ for both silicates. HD (D$_2$) molecules form via D atom abstraction of chemisorbed H (D) atoms from OH (OD) groups, with a cross section of $\sigma _{f,D_2}$  = 7.0$\times10^{-18}$ cm$^{2}$. Quantum chemical calculations on enstatite and forsterite amorphous nano-clusters were used to analyze chemisorption and abstraction energies of H atoms. The formation of OH groups on forsterite is barrierless, while enstatite has a small energy barrier. H$_2$ abstraction from hydroxyl groups is barrierless in both silicates. The results support our interpretation of IR spectral changes during D atom exposure as addition and exchange reactions, with catalytic molecular deuterium formation. These findings, obtained at 300 K, are relevant to astrophysical environments like photodissociation regions and interstellar clouds at lower temperatures. Amorphous silicates, like carbon grains, undergo hydrogenation in the diffuse interstellar medium due to interactions with atomic hydrogen and UV photons. The detection of these components on comet 67P/Churyumov-Gerasimenko connects primitive solar system objects to interstellar dust, offering new insights into their evolution.

\end{abstract}


\section{Introduction} \label{sec:intro}

Molecular hydrogen is the most abundant molecule in the Universe. It plays a critical role in astrochemistry by contributing directly or indirectly to all reactions involved in the formation of both simple and complex molecules within the interstellar medium (ISM) \citep{williams1999}. Additionally, H$_2$ is vital for the ISM's physical and chemical properties, acting as a cooling and heating agent and triggering the collapse of molecular clouds that lead to star formation. This prompts considerable interest in understanding the processes behind the formation of H$_2$ in various space environments. It has long been acknowledged that, under ISM conditions, H$_2$ cannot be efficiently formed in the gas phase to account for its abundance \citep{gould1963}. Instead, it is firmly established that H$_2$ forms through catalytic reactions occurring on the surfaces of interstellar dust grains, which consist mainly of silicates and carbonaceous materials \citep{hollenbach1970}.

When the grain temperatures are below 20 K, interactions with H atoms from the gas phase are primarily governed by weak adsorption forces. In these cold conditions, H atoms are mobile, allowing them to diffuse across the grain surface and recombine to form H$_2$, primarily through the Langmuir-Hinshelwood (LH) mechanism \citep[\& references therein]{cazaux2004}. Additionally, gas-phase H atoms can directly interact with an already adsorbed H atom or with the grain surface near where an H atom is adsorbed, leading to H$_2$ formation via the Eley-Rideal (ER) or the Harris-Kasemo (H-K) “hot-atom" mechanism. These mechanisms become relevant only when there is sufficient coverage of H atoms on the grain surface. Experimental studies of H$_2$ formation at these low temperatures have used refractory silicate grain analogs, including polycrystalline and amorphous silicates \citep[e.g.][]{Pirronello1997a, Pirronello1997b, perets2007, vidali2007, he2011, gavilan2012}, as well as carbon grain analogs like amorphous carbon and graphite \citep[e.g.][]{Pirronello1999, islam2007, latimer2008}. This process has also been explored on amorphous water ice surfaces \citep{Manico2001, Hornekaer2003}. Similarly, theoretical studies have examined the recombination of physisorbed H atoms on carbon-based graphite surfaces \citep[e.g.][]{farebrother2000, morisset2004} and water ice \citep[e.g.][]{takahashi1999, al2007}.
 
At high grain temperatures ($>$ 20 K) typical of environments like photo-dissociation regions (PDRs) and the hot cores/corinos of young stellar objects, H atom recombination mechanisms involving physisorption sites are not efficient. This is because H atoms desorb quickly from the grain surface, preventing recombination. Laboratory experiments have shown that molecular H$_2$ formation can indeed take place on hydrogenated carbon grains at room temperature, via the ER mechanism on chemisorbed H atoms of CH aliphatic bonds \citep{Mennella2008, Mennella2011a}. This was the first experimental evidence for the catalytic production of hydrogen and demonstrated that chemisorption sites are essential for the process to occur at high grain temperatures. Similar results were obtained on the aliphatic C-H sites of a superhydrogenated polycyclic aromatic hydrocarbon molecule coronene at 300 K \citep{mennella2011b, thrower2012}. Subsequently, theoretical studies of H$_2$ formation via different mechanisms have been extended to include both physisorption and chemisorption of H atoms and such studies were primarily performed on silicate surface models \citep[e.g.][]{goumans2009, navarro2014, kerkeni2013}. For a concise summary of H$_2$ formation on refractory grains, refer to \citet{suhasaria2021}. For a comprehensive review encompassing observational, experimental, and theoretical aspects of H$_2$ formation in its entirety, see \citet{wakelam2017}.
 
This study extends the laboratory investigation of H$_2$ formation at elevated grain temperatures to include hydroxylated Mg-rich amorphous silicate grains. In the ISM, amorphous silicates constitute a higher proportion ($\sim$98\%) compared to their crystalline forms, with the silicates predominantly being Mg-rich \citep[e.g.][]{kelley2017}. Our experimental investigation presents, for the first time, evidence of the catalytic role played by hydroxylated silicate grain analogs of various compositions in facilitating H$_2$ formation at grain and H atom temperature of 300 K, a condition relevant to astrophysical environments such as PDRs and maintain their validity at lower temperatures of interstellar clouds. 

\section{Experiments and Results} \label{sec:ER}
The Mg-rich amorphous silicates considered in the present research as interstellar dust analogs have been reported in detail in previous work \citep{mennella2020}; only a brief description of the salient aspects of the preparation of the samples and of their morphological, chemical and spectral properties is given here. The samples were synthesized by laser ablation of targets of oxides mixtures and subsequent condensation of vapor on ZnSe windows. We adopted a pulsed Nd-YAG solid-state laser at 1024 nm, along with its second and fourth harmonics. Pellets (13 mm in diameter) of mixtures of periclase (MgO) and quartz (SiO$_2$) at the stoichiometric ratio of enstatite (1:1) and forsterite (2:1) were used as target. The resulting samples have an Mg/Si atomic ratio of 1.1 for SIL1 (enstatite type) and 1.7 for SIL2 (forsterite type), as determined by Electron Dispersive X-ray (EDX). The Field Emission Scanning Electron Microscope (FESEM) images show that mainly a fluffy structure of amorphous grain aggregates characterizes the morphology of the silicate samples \citep{rotundi2000}.
  
\begin{figure}[t!]
\centering
    \resizebox{\hsize}{!}{\includegraphics{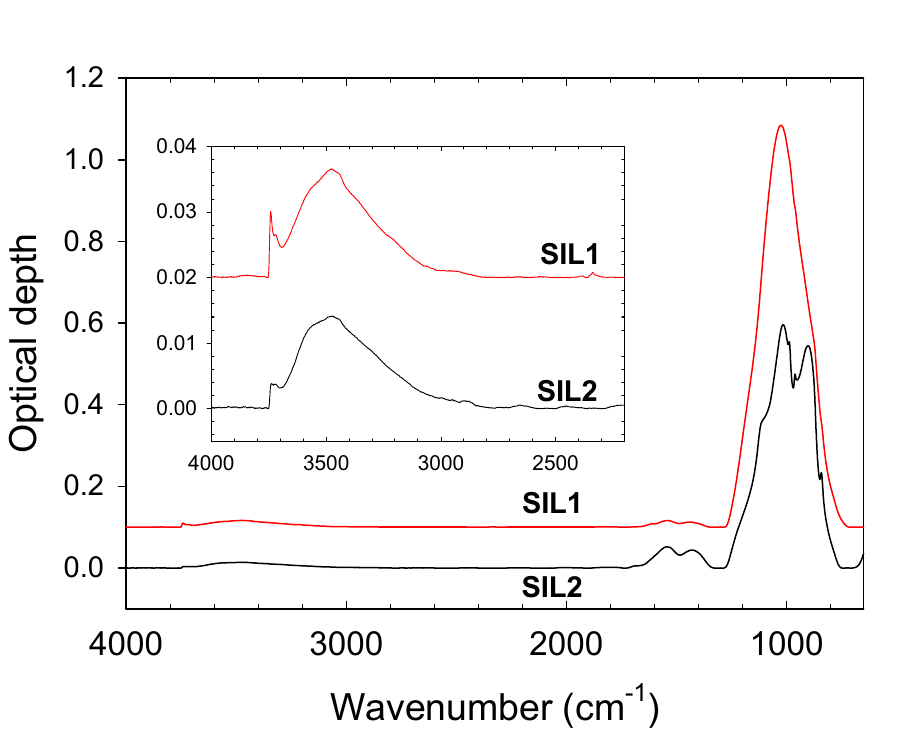}}
\caption{Continuum subtracted mid-IR spectra of SIL1 and SIL2. The details of the O-H stretching region are shown in the inset. The spectra are offset in ordinate for the sake of clarity. \label{fig1}}
\end{figure}

Figure \ref{fig1} shows the optical depth of the silicates measured with a Fourier Transform Infrared (FTIR) spectrometer (Bruker Vertex 80v). Transmittance spectra were acquired at room temperature, in the range 4000 -- 700 cm$^{-1}$ at a spectral resolution of 2 cm$^{-1}$. A strong band around 1025 cm$^{-1}$ after baseline correction characterizes the Si-O stretching region of SIL1. As the Mg/Si ratio increases for SIL2, this band shifts to 1015 cm$^{-1}$,  a second strong band at 903 and three weak features at 988, 961, and 841 cm$^{-1}$ appear, along with a shoulder at 1105 and cm$^{-1}$.  The peak positions and relative intensities of these six features closely align with the bands reported for synthetic microcrystalline forsterite by \citet{jager1998}. According to their study, all the bands correspond to asymmetric stretching modes of SiO$_4$, except for the feature at 841 cm$^{-1}$, which is attributed to the symmetric stretching mode of SiO$_4$. Interestingly, this result appears to indicate a degree of crystallinity in  SIL2  that requires further investigation to better understand its origin. The bands at 1542 and 1431 cm$^{-1}$, more intense in the case of SIL2, caused by traces of magnesium carbonate, are typical artifacts of amorphous silicate preparation with sol-gel and laser ablation methods \citep{jager2003, ciaravella2018, rouille2024}. 

A broad band centered at 3474 cm$^{-1}$, produced by adsorbed water and chemisorbed hydroxyl groups, possibly due to ambient contamination during sample preparation and handling characterizes the O-H stretching region. Superimposed to this band is a doublet with sharp features at 3741 and 3716 cm$^{-1}$. Bands falling around 3700 cm$^{-1}$ have been suggested to arise due to isolated Si-OH and Mg-OH groups \citep[and references therein]{jager2003}. However, the intensity decrease of the doublet with increasing the Mg/Si ratio from SIL1 to SIL2 should exclude their attribution to Mg-OH groups in our samples. Therefore, we can confidently attribute the high and low frequency peaks of the doublet to isolated and terminal silanol groups, respectively, following the assignment of bands at 3745 and 3715 cm$^{-1}$ by \citet{gallas2009} in silica gel and mesoporous silica.

\begin{figure}[t!]
\centering
    \resizebox{\hsize}{!}{\includegraphics{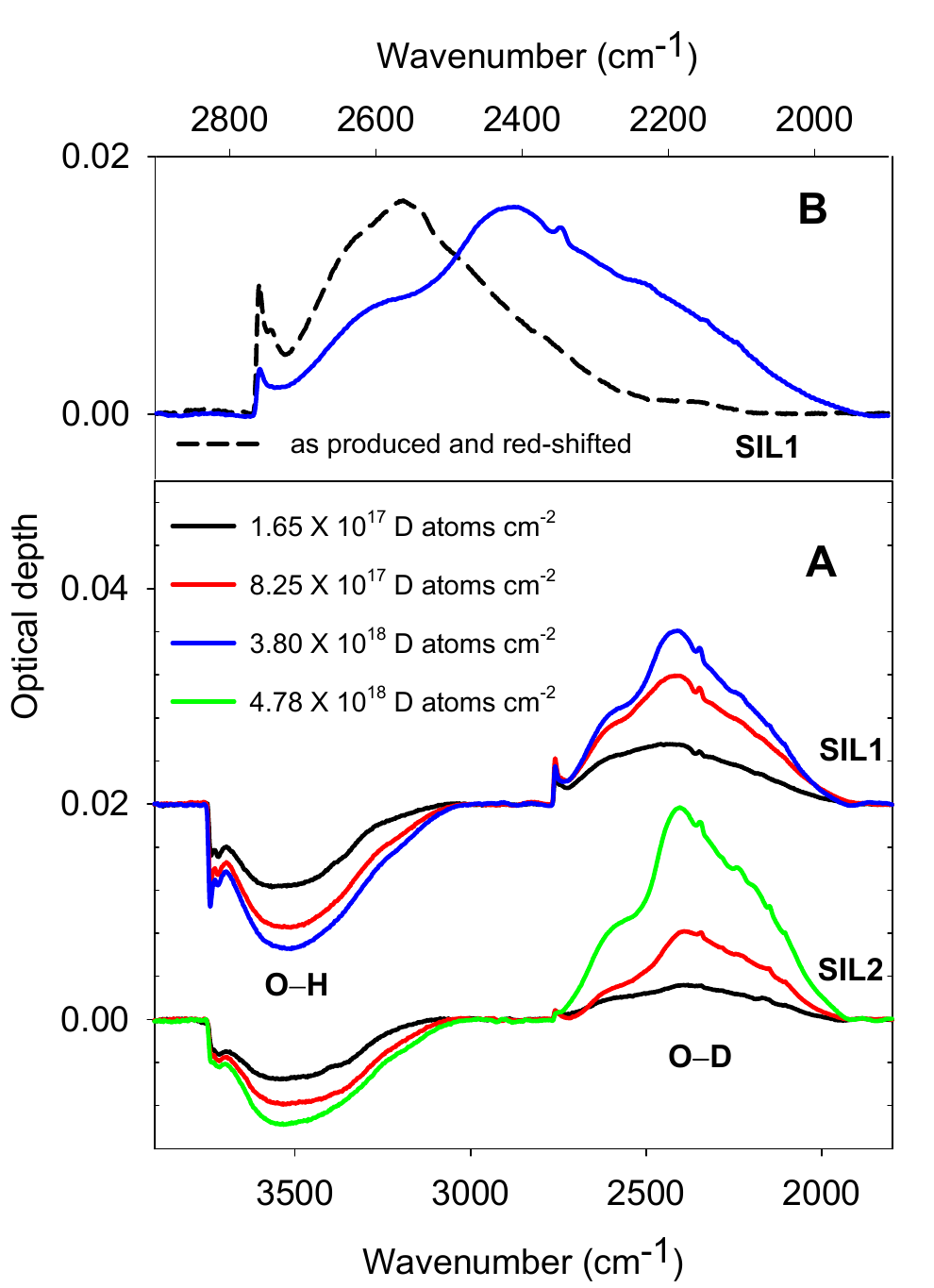}}
\caption{A) Evolution of the O-H and O-D stretching bands during D atom irradiation of the SIL1 and SIL2 silicates. The spectra  of the exposed silicates are shown after subtraction 
of the spectrum of the samples as produced. The SIL1 spectra are offset in ordinate for the sake of clarity.
B) The O-D stretching band of SIL1 after D exposure compared with the O-H mode of SIL1 as shown in the inset to Figure \ref{fig1}. The later has been red shifted, reducing the frequency of a factor 1.36, to take into account the H/D isotope substitution.
 \label{fig2}}
\end{figure}

To investigate the interaction of hydrogen with amorphous silicates we exposed the two silicate samples to deuterium to discriminate their effects from those of possible reactions involving hydrogen atoms already present on sample surface. An atomic deuterium beam, with D atoms having a Maxwellian velocity distribution at 300 K \citep{mennella2006}, have been produced by microwave dissociation of D$_2$. The samples were placed on a rotating sample holder at the right angle intersection between the atomic beam and the IR beam of the FTIR spectrophotometer inside a vacuum chamber with a base pressure of 2$\times$10$^{-8}$ mbar. After D atom exposure, the sample was rotated by 90$\degr$ to face the IR beam, and we measured the IR spectrum. Figure \ref{fig2}A shows the changes in the O-H and O-D stretching spectral regions induced in amorphous silicates as a function of D atom fluences. The latter were derived from the D atom exposure times using an atomic flux of 5.5$\times$10$^{14}$  atoms cm$^{-2}$ sec$^{-1}$. For the experimental conditions of the present experiment, D$_2$ dissociation is 0.66 \citep{mennella2002}. The activation of a broad band centered at 2414 cm$^{-1}$ with a shoulder at 2580 cm$^{-1}$ and a sharp doublet at 2760 and 2742 cm$^{-1}$ takes place in both types of silicate. At shorter wavelengths, an absorbance decrease in the exposed sample (resulting in a negative optical depth difference in Figure \ref{fig2}A) centered at 3520 cm$^{-1}$ (broad band) and bands at 3741 and 3716 cm$^{-1}$ (sharp doublet) are observed. This behavior indicates that D atoms induced a depletion of the O-H bonds already present in the starting silicate samples and resulted in the formation of new O-D bonds, whose stretching modes shift to lower wavenumbers due to the isotope substitution. To evaluate the frequency shift for our spectra we have considered the sharp peaks of the SiOH and SiOD groups at 3741 and 2760 cm$^{-1}$, respectively. The measured ratio of 1.36 for the peak positions is in good agreement with the expected shift due to the variation of the reduced mass of 1.37 for the O-H and O-D groups. Taking into account this value one can see that the shoulder at 2580 cm$^{-1}$ corresponds to the 3520 cm$^{-1}$ band. Moreover, the whole feature centred at 2414 cm$^{-1}$ is broader than the OH stretching feature present before D atom exposure (see Figure \ref{fig2}B). The relative intensities of the features at 2580 and 2414 cm$^{-1}$ change during D exposure, with the latter increasing compared to the high frequency component as fluence of D atoms, F$_D$, increases.

\begin{figure}[ht!]
\centering
    \resizebox{\hsize}{!}{\includegraphics{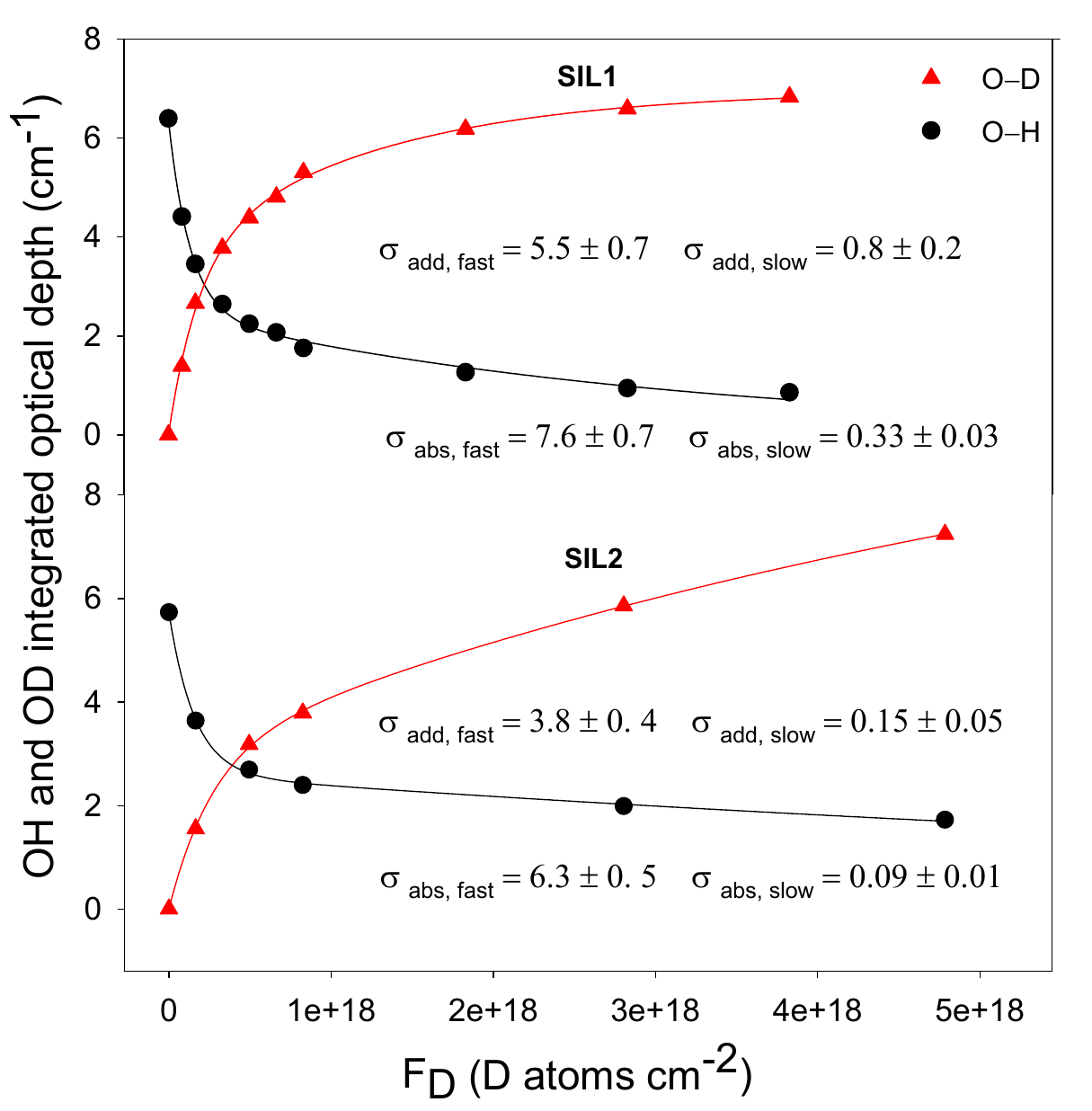}}
\caption{ Evolution with D atom fluence of the O-H and O-D integrated band optical depth during D atom exposure of SIL1 and SIL2 silicate grains. The best fit to the data of the
relations $a_{1}[1-\exp (-\protect\sigma _{add,fast}F_{D})]+a_{2}[1-\exp (-%
\protect\sigma _{add,slow}F_{D})]$ and $b_{1}\exp (-\protect\sigma %
_{abs,fast}F_{D})+b_{2}\exp (-\protect\sigma _{abs,slow}F_{D})$, respectively, for the
addition and abstraction processes, are also shown. The estimated cross sections are expressed in units of 10$^{-18}$ cm $^{2}$.
\label{fig3}}
\end{figure}

The evolution of the integrated optical depth of the O-H and O-D stretching bands with $F_{D}$ is shown in Figure \ref{fig3}. The behaviour is similar to that found for hydrogenated carbon grains exposed to D atoms. As in that case, it can be interpreted in terms of an H/D exchange reaction occurring through the two step sequence: 1) dehydrogenation of an OH group via abstraction of hydrogen by a D atom  (with the formation of an HD molecule) and 2) formation of a OD group via D addition on the dangling bonds formed in the previous step when a second atom impinges on the sample. In addition to the OD groups resulting from exchange reactions, further deuteroxyl groups form during D exposure, see next section.  It is worth mentioning that the interaction of D atoms with OD groups generates D\textsubscript{2} molecules as a result of the D/D exchange but does not induce spectral changes. The interaction constantly converts atomic hydrogen into molecular hydrogen, whereas the surface of hydroxylated amorphous silicates remains unaltered.

 The trend of the integrated optical depth of the O-H and  O-D bands with $F_{D}$ has been used to estimate the abstraction and addition cross-sections. The experimental data are well fitted with an initial fast decay rate, with a cross section $\sigma _{fast}$ and a slow rate with $\sigma _{slow}$  rather than a single cross section. This behavior has also been observed in hydrogenated amorphous carbon films and grains and interpreted as related to the heterogeneous nature of the materials that have hydrogen sites that are difficult to interact with D atoms \citep[e.g.][]{kuppers1995, Mennella2008}. Similar morphology of the samples, characterized by a fluffy aggregation of small grains with an amorphous structure, despite their completely different chemical compositions, may help explain the variations in decay rates. The best fits values to the data are shown in Figure \ref{fig3}. The values of $\sigma _{abs, fast}$ and $\sigma _{add, fast}$ for the two experiments are comparable within the errors, indicating  that deuteroxylation and molecular hydrogen production does not depend on the chemical composition (Mg/Si ratio) of the amorphous silicates under consideration. The average values $\sigma _{f,OD}$  = 4.2$\times10^{-18}$ cm$^{2}$ and $\sigma _{f,D_2}$  = 7.0$\times10^{-18}$ cm$^{2}$ can be considered as representative of hydroxyl groups and molecular hydrogen formation in hydroxylated Mg-rich amorphous silicates. The abstraction cross sections obtained in the present work compare well with those reported for hydrogenated amorphous carbon grains and films, respectively, 3$\times10^{-18}$ and 5 $\times10^{-18}$ cm$^{2}$ \citep{biener1995, Mennella2008}. As in those cases, the small value of the cross section is indicative of an ER process driving the catalytic formation of HD (D$_2$) molecules on Mg-rich amorphous silicates.    

\section{Discussion} \label{sec:DC}
The use of D atoms allows us to identify
 two main effects of their interaction with Mg-rich amorphous silicates. As indicated by the spectral variations of Figure \ref{fig2}, the optical depth decrease at 3520 cm$^{-1}$ (broad band) and at 3741 and 3716 cm$^{-1}$ (sharp doublet) indicates a reduction in the number of OH bonds that were already present in the samples. These spectral changes are evidence for the first step of the exchange reaction sequence: abstraction of H atom of a hydroxyl group by an impinging D atom producing an HD molecule. Abstraction leaves a site available to form an OD group when a second atom arrives (addition reaction). This is the second step of the sequence and it is testified by the appearance in the OD spectral region of a shoulder at 2580 cm$^{-1}$ and a of sharp doublet at 2760 and 2742 cm$^{-1}$ in Figure \ref{fig2}. On the other hand, the increase of the optical depth at frequency lower than 2414 cm$^{-1}$ corresponds to the formation of OD groups that are not the result of exchange reactions with H  atoms (see Figure \ref{fig2}B). These groups are activated by D atoms in defects on amorphous silicates, which determine a wide range of physisorption and chemisorption binding sites and give rise to multiple and independent absorption overlapping \citep{mennella2020}. As a result, the deuteroxyl band width in the current experiments is quite broad. Additionally, it was previously thought that the hydroxylation of silicate could only occur through the interaction with high-energy protons in the keV to MeV range \citep[e.g.]{zeller1966, djouadi2011, ichimura2012, schaible2014}. However, the present experiment demonstrates that the same process can be achieved using H atoms with extremely lower energies (a few tenths of meV).

\begin{figure}[t!]
\centering
    \resizebox{\hsize}{!}{\includegraphics{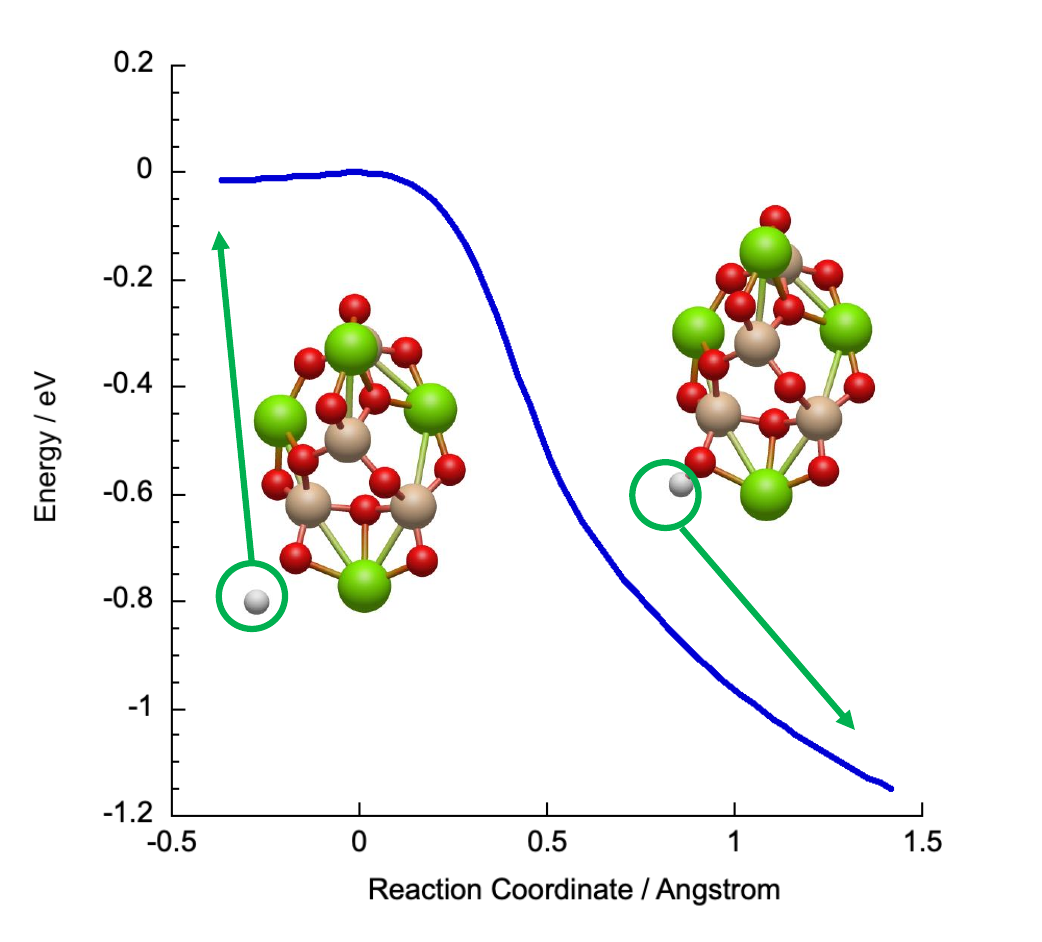}}
\caption{The potential energy curve for the chemisorption of a H atom on an O atom of the enstatite nanocluster, calculated using the M062X/cc-pVTZ methodology. The atom colors are as follows: white for hydrogen (H), red for oxygen (O), beige for silicon (Si), and green for magnesium (Mg). Upon OH formation the O-Mg bond elongates from 1.8 to 2.2\AA.  \label{fig4}}
\end{figure}

Quantum chemical computations on nano silicate clusters have been performed to analyse the interaction of H atoms with our amorphous silicates. The inherently non-crystalline structure of the stable nanosilicate cluster model naturally simulates the effects of grain amorphicity \citep{oueslati2015}. Specifically, we performed density functional theory (DFT) calculations by means of the hybrid-meta GGA functional M06-2X \citep{m06} and the cc-pVTZ basis set \citep{tz} of impinging H atoms on two nanocluster models: enstatite (Mg$_4$Si$_4$O$_{12}$) for SIL1 and forsterite ((MgO)$_6$(SiO$_2$)$_3$) for SIL2, the details of the structures are reported elsewhere \citep{kerkeni2013, kerkeni2019}. We allowed all the atoms in the nanocluster to relax during optimization and found a H-physisorption well of 0.15 eV (64 meV with ZPE) on enstatite. On the other hand, chemisorption of an incoming hydrogen atom proceeds with a relatively low energy barrier for the formation of an OH group--13 meV (3 meV with Zero Point Energy (ZPE)) and is highly exothermic, releasing 1.7 eV (1.4 eV with ZPE) (see Figure \ref{fig4}). The addition of a second gas-phase H atom to form another OH group occurs without an energy barrier and the process is also exothermic. Note that \cite{kerkeni2017} found that there is also no barrier to the second H atom binding to a Mg site close to an already formed OH group . For forsterite, the H-physisorption energy well is 0.11 eV (60 meV with ZPE). No energy barriers were identified for H atom chemisorption, which is highly exothermic, releasing 1.83 eV (1.6 eV with ZPE). This finding remains consistent across various oxygen sites investigated.

The next step of our analysis focused on investigating H$_2$ formation through the abstraction of a chemisorbed H atom from a hydroxyl group in the enstatite nanocluster through the ER mechanism. Our computations show that H$_2$ formation through this abstraction proceeds without an energy barrier in both nanoclusters. Additionally, the other significant result is the overall exothermicity of 4.1 eV for the abstraction reaction, indicating that the entire reaction mechanism is energetically favorable and likely to occur readily (see the Appendix
\ref{sec:X} for further details).

The nano cluster model we employed has proven useful in demonstrating that both the addition and abstraction of a hydrogen atom on amorphous Mg-rich silicates can occur without an energy barrier. However, after hydrogenation, the model has only bridging silanol configuration which is an oversimplified representation of the bonding configuration in our samples. In fact, the experimentally observed IR peaks are quite broad indicating an overlap of multiple OH stretching modes due to different bonding configurations possibly connected through both hydrogen and covalent bonds. The model cannot reproduce the complexity of the experimental infrared spectrum.

Our interpretation of the IR spectral variations in the experiment is robust and aligns with the spectral evolution observed upon exposure of hydrogenated carbon grains to D atoms. Nonetheless, other reactions could occur during the irradiation of hydroxylated silicates by D atoms. For example, if vicinal silanol groups are present, they could interact due to the energy released during the formation of SiOD, leading to the generation of H$_2$O molecules. Furthermore, D$_2$ could also form on MgD groups through ER reactions with D atoms. However, the examination of these reactions using DFT calculations goes beyond the scope of this study.

\section{Astrophysical implications} 
Silicates play a significant role in the evolutionary process that transports dust grains  from  their circumstellar formation sites to diffuse and dense interstellar clouds, which are the birth places of stars, disks, and planetary systems \citep{henning2010}. In this process they can be hydoxylated in their formation stage and in the ISM. In fact, in circumstellar shells of evolved stars, during the formation of olivine and pyroxen type silicates, the presence of water can give rise to hydoxyl groups as well as to metal-OH isolated groups \citep{gail1998, jager2003}. Silicate grains can also get hydroxylated by H atoms in the interstellar clouds  and in PDRs and act as catalysts for molecular H$_2$ formation.  The cross sections estimated in the current experiments, carried out on samples at 300 K with D atoms also at 300 K, are relevant for predicting the formation of hydroxyl groups and H$_2$ in these regions where H atoms and grain temperatures typically range from 100 to 600 K and from 10 to 50 K, respectively \citep[e.g.][]{sorrell1990, habart2004}. In fact as indicated by DFT computations, the formation  OH groups on silicates through H atom addition takes place without a significant energy barrier, suggesting no dependence of the cross section on gas temperature. Concerning variations of the formation cross section on grain temperature, only a minor reduction  is expected at low temperatures. For example, when carbon grains are exposed to atomic H, the cross section for the formation of aliphatic C-H bonds decreases by a factor of about 3 as the grain temperature decreased from 300 to 12 K \citep{Mennella2008}. Note that for grain temperatures lower than 20 K, H atom recombination on grain surfaces via LH and/or HK mechanisms can also be active and contribute to H$_2$ formation. 

 \begin{figure}[t!]
\centering
    \resizebox{\hsize}{!}{\includegraphics{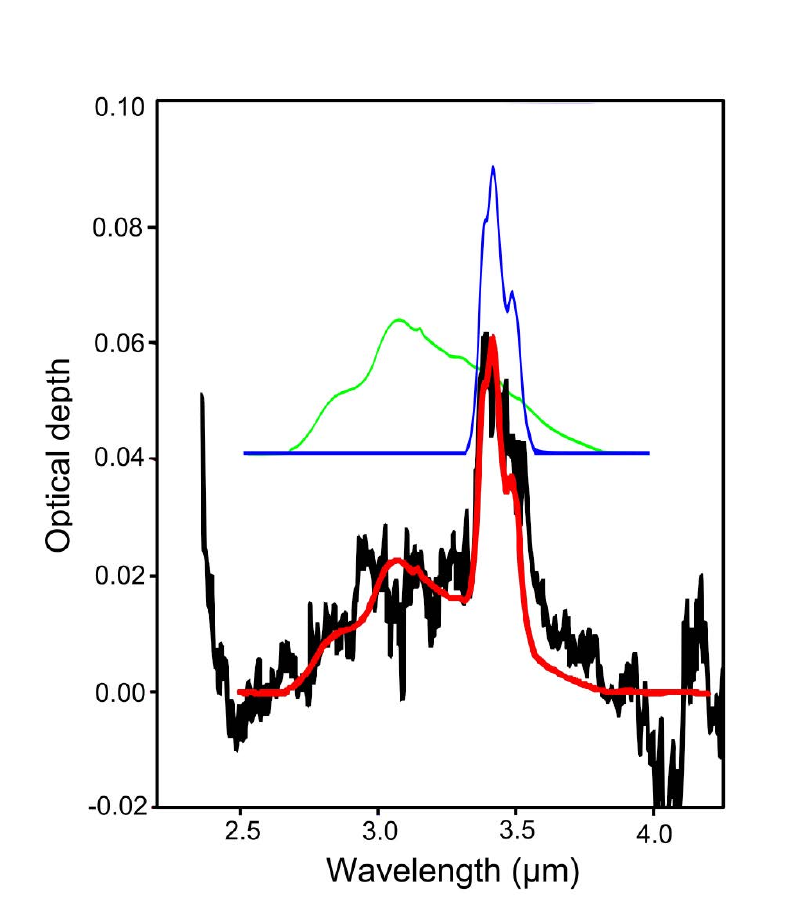}}
\caption{Comparison of a linear combination (red line) of optical depth of hydroxylated silicates SIL2 (green line) and hydrogenated amorphous carbon grains (blue line) \citep{mennella2002} with the average of three ISO spectra of Cyg. OB2 12 (black line). The astronomical spectrum is adapted form Extended Data Fig. 3 by \citet{potapov2021}. The hydroxylated silicate SIL2 was obtained by blue-shifting the deuteroxylated silicate spectrum, applying a factor of 1.36 for the D/H isotope substitution. The individual components are offset in ordinate.
\label{fig5}}
\end{figure}

An accurate prediction of the level of hydroxylation of silicates in space must also take into account the processes that lead to the OH group destruction, which is driven by UV photons in interstellar clouds and PDRs. To our knowledge, no experimental determination of the destruction cross section of OH bonds  on silicates has been done. However, since the typical bond dissociation energies for O-H and C-H bonds differ by a few tenths of eV \citep{sanderson1976}, one expects comparable photo-dissociation cross sections by Ly-$\alpha$ photons. As a first approximation, the C-H bond destruction cross section of  1$\times10^{-19}$ cm$^{2}$  evaluated in Ly-$\alpha$ irradiation experiments of hydrogenated carbon grains can be used as reference value along with the formation cross section of OH bonds estimated in this work, to predict whether or not silicates are hydroxylated for interstellar environments whose UV and H atom fluxes are known. To this end, we can apply the first order kinetic model outlined to describe the  evolution of C-H bonds under the contrasting action of UV photons and hydrogen atoms, see \citet{mennella2002, mennella2003} for details. As an example we consider silicate grains injected from circumstellar shell into diffuse interstellar medium (DISM hereafter), which are characterised by a UV flux $\Phi_{UV}$ = 8$\times10^7$ photons s$^{-1}$ cm$^{-2}$ and an atomic H flux $\Phi_{H}$ = 8$\times10^6$ atoms s$^{-1}$ cm$^{-2}$  \citep{mathis1983, sorrell1990, mennella2003}. 
Considering the OH group cross sections for formation via H atom addition and destruction by UV photons mentioned earlier, the corresponding rates of formation and destruction are, respectively, $R_{f, OH}$ = $\sigma _{f,OD} \Phi_{H}$ = 3.4$ \times10^{-11} $ s$^{-1}$ and $R_{d, OH}$ = $\sigma _{d,UV} \Phi_{UV}$= 8$\times10^{-12} $ s$^{-1}$.  Regardless of its initial value, the steady state level for silicate hydroxylation, normalized to its maximum value in the considered silicates is (refer to eq 18 in \citet{mennella2002}: $n_{OH}/n_{OH, max}$ = $R_{f, OH}$/($R_{f, OH}$ + $R_{d, OH}$) = 0.81. This value is attained in approximately $10^3$ years, which is an extremely brief duration when compared to the cloud life time of 3$\times10^7$ yrs. This result suggests that competition between the two opposing processes examined  leads to the presence of hydroxylated silicates in DISM.  Currently, we do not have information on their quantity per hydrogen column density; additional  work is necessary to assess this quantity. Under these circumstances, the formation of molecular H$_2$ occurs at a rate ${R_{H_2}}$ = $\sigma _{f,D_2} \Phi_{H}$ = 5.6$\times10^{-11} $ s$^{-1}$, taking into account the experimental value of $\sigma _{f,D_2}$ derived in this work.

The spectrum of hydroxylated silicates SIL2 exposed to 8.25 $\times 10^{17}$ D atoms $cm^{-2}$ and a DISM ISO spectrum towards Cyg OB2 12 by \citet{potapov2021} are compared in Figure \ref{fig5}; the figure also includes the laboratory profile of hydrogenated amorphous carbon grains to mimic the C-H stretching aliphatic 3.4 $\mu$m micron band. Combining the optical depth of two laboratory spectra linearly result in a successful spectral match. The close spectral agreement suggests that hydroxylated silicates and hydrogenated carbon grains are the primary carriers of the interstellar spectrum. An interesting aspect of this interpretation is that both analogues are consistent with the physical conditions of the DISM. In fact, they are the result of laboratory experiments that simulate grain processing, including the interaction with hydrogen atoms and UV photons, which drives the evolution of dust in the DISM. Note that while the reproduction of the O-H spectral features is only qualitative (as discussed earlier), about 65 ppm of carbon relative to hydrogen are locked into hydrogenated carbon grains, which are responsible for the interstellar aliphatic feature. 

\citet{potapov2021} associated the O-H region of the spectrum towards Cyg OB2 12 with water molecules physisorbed (trapped) on interstellar silicate grain surface. However, a subsequent study by the same authors found that  water  on silicates cannot survive in DISM \citep{potapov2024}.
The O-H region of the spectrum toward Cyg OB2 12 can also be associated with water ice. In fact, it is understood that water molecules form through reactions between hydrogen and oxygen (mainly atomic in diffuse environments) on grain surfaces, creating ice at the low interstellar temperatures. However, modeling  of the growth of ice mantles on interstellar grains has shown that in the DISM, the thickness of a water ice mantle is limited to no more than one monolayer due to the high efficiency of UV photodesorption \citep{cuppen2007}.

Water ice formation has also been investigated in numerous laboratories using different simulated grain surfaces under conditions mimicking the DISM. For this discussion, we focus on the case of water ice formation on amorphous silicates, as studied by \citet{jing2011}. The authors conducted a series of experiments involving simultaneous and sequential deposition of deuterium and oxygen (in atomic and/or molecular form) at comparable fluxes onto amorphous silicates at temperatures of 15–25 K. Water ice formation was observed in all cases, though with varying efficiency, except when a hydrogen-rich surface was present. This scenario resembles DISM conditions, where the flux of hydrogen atoms is much higher than that of oxygen atoms (by a factor of $\sim$ 10$^4$). In such environments, the interaction of silicates with atomic hydrogen dominates, making silicate hydroxylation more likely than water ice formation.

Hydroxylated silicates can form in situ in the DISM and can also be present on the surface of comet 67P/Churyumov-Gerasimenko \citep{mennella2020}. The same applies to hydrogenated amorphous carbon grains \citep{Raponi_2020}. As suggested by laboratory simulation of the processes driving the evolution of dust in space, the observation of these two primary refractory grain components in different environments suggests an evolutionary link between dust in the ISM and primitive objects of the solar system.
 
\section{Conclusions} \label{sec:C}
There is a striking similarity between the present results and the evolution with D exposure of the aliphatic C-H bonds in hydrogenated carbon grains (see Figures 1 and 2 of \citet{Mennella2008}). Taken as a whole these experimental results show that, unlike recombination on physisorption sites which is inefficient for grain temperature greater than 20 K, molecular hydrogen formation on chemisorption sites of hydrogenated amorphous carbon and hydroxylated amorphous silicates, the primary refractory components of the cosmic dust, can easily proceed from low to high grain temperatures. Another noteworthy aspect is the fact that the production occurs in a catalytic manner, i.e. continuously, leaving the grains in the same state they were in before the interaction with hydrogen atoms. These results point the way towards a complete solution of the problem of molecular hydrogen formation at high temperatures.

Finally, we have shown that Mg-rich amorphous silicates, like amorphous carbon grains, undergo hydroxylation/hydrogenation in the DISM, with signatures of both observed on comet 67P/Churyumov-Gerasimenko. This finding supports the evolutionary connection between interstellar dust in DISM and primitive solar system objects.

\acknowledgments
This research has been funded by the Istituto Nazionale di Astrofisica (INAF). TS acknowledges the support of the European Research Council through the ERC Advanced Grant Origins 832428 under the Horizon 2020 Framework Program. The authors also express their gratitude to the Deanship of Graduate Studies and Scientific Research at Taif University for their financial support of this work.

\appendix

\section{DFT Calculations of the Energetics for Eley-Rideal Abstraction of H$_2$}\label{sec:X}

\begin{figure*}[ht!]
\centering
    \resizebox{\hsize}{!}{\includegraphics{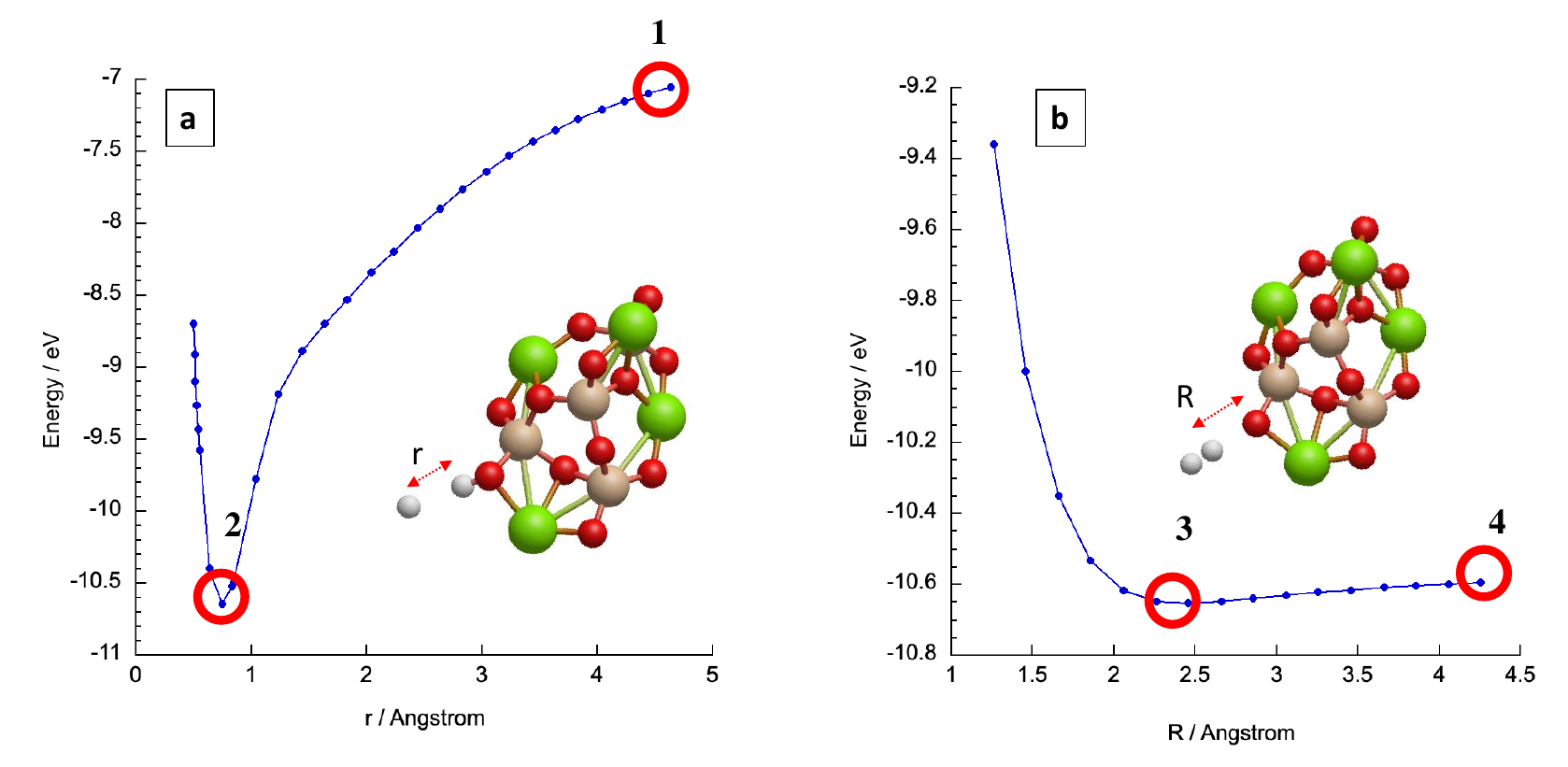}}
\caption{The ER potential energy surface along the minimum energy path for (a) H atom approaching an hydroxyl group from infinity and (b) H$_2$ molecule abstraction from the enstatite nanocluster.\label{fig6}}
\end{figure*}

The cuts of the two-dimensional ER potential energy surface along the minimum energy path for two cases are presented in Figure \ref{fig6}. The left panel depicts an H atom approach from a far distance (red circle marked with 1) to an hydroxyl group of the enstatite nanocluster, while the right panel illustrates the abstraction of an H$_2$ molecule from the O atom of the nanocluster to the gas phase. Following \citet{farebrother2000}, we constrained the approach of H atom and the release of H$_2$ molecule along the line of OH bond to keep the problem to two dimensions (R, r). The approach of an H atom is barrierless, following a minimum energy path (red circle marked with 2). This occurs when the distance between the incoming hydrogen atom and the hydrogen atom of the hydroxyl group (r) is 0.75 \AA, signaling H$_2$ formation, with distance between H$_2$ molecule and O atom of the nanocluster (R) being 2.3 \AA. The reaction is exothermic by 3.7 eV. The cross-over between the reaction coordinates (r and R) in the potential energy surface occurs at an energy of -10.65 eV (indicated by the red circles 2 in Figure \ref{fig6}a and 3 in Figure \ref{fig6}b), forming a van der Waals complex between H$_2$ and the nanocluster. The release of the H$_2$ molecule from point 3 into the gas phase is slightly endothermic by 0.07 eV, which reduces the overall exothermicity by this small amount. Overall, the reaction mechanism is energetically favorable and proceeds easily.

\bibliography{References}{}
\bibliographystyle{aasjournal}

\end{document}